\documentclass[conference,compsoc]{IEEEtran}
\ifCLASSOPTIONcompsoc
  \usepackage[nocompress]{cite}
\else
  \usepackage{cite}
\fi
\ifCLASSINFOpdf
  \usepackage[pdftex]{graphicx}
\else
\fi
\begin{document}
%
\title{Decentralized AI: Permissionless LLM Inference on POKT Network}

\author{\IEEEauthorblockN{Daniel Olshansky}
\IEEEauthorblockA{Grove Inc.}
\and
\IEEEauthorblockN{Ramiro Rodr\'iguez Colmeiro}
\IEEEauthorblockA{Pocket Scan Technologies LLC}
\and
\IEEEauthorblockN{Bowen Li}
\IEEEauthorblockA{POKT Network Foundation}}


%


\maketitle

\begin{abstract}
POKT Network's decentralized Remote Procedure Call (RPC) infrastructure, surpassing 740 billion requests since launching on MainNet in 2020, is well-positioned to extend into providing AI inference services with minimal design or implementation modifications. This litepaper illustrates how the network's open-source and permissionless design aligns incentives among model researchers, hardware operators, API providers and users whom we term model Sources, Suppliers, Gateways and Applications respectively. Through its Relay Mining algorithm, POKT creates a transparent marketplace where costs and earnings directly reflect cryptographically verified usage. This decentralized framework offers large model AI researchers a new avenue to disseminate their work and generate revenue without the complexities of maintaining infrastructure or building end-user products. Supply scales naturally with demand, as evidenced in recent years and the protocol's free market dynamics. POKT Gateways facilitate network growth, evolution, adoption, and quality by acting as application-facing load balancers, providing value-added features without managing LLM nodes directly. This vertically decoupled network, battle tested over several years, is set up to accelerate the adoption, operation, innovation and financialization of open-source models. It is the first mature permissionless network whose quality of service competes with centralized entities set up to provide application grade inference.
\end{abstract}


%
\IEEEpeerreviewmaketitle

\section{Introduction}

\subsection{ LLM Inference \& Web3 Full Nodes}
The advent of OpenAI’s ChatGPT brought foundation models into the mainstream. With it, the ecosystem of fine-tuning, distributing, evaluating and optimizing models has become ubiquitous. Companies like Meta are training and open-sourcing~\cite{metaIntroducingMeta} models ranging from 8B (small) to over 400B (large) parameters, often referred to as Language Models (LMs), Large Language Models (LLMs), or Large Multimodal Models (LMMs). Platforms like HuggingFace have become central hubs for sharing and discovering new models, hosting hundreds of thousands~\cite{greataipromptsEveryHugging} of open-source models from institutions and independent researchers.

Although some models can be hosted on personal devices~\cite{pytorchExecuTorchAlpha}, most AI engineers~\cite{latentRiseEngineer} rely on third-party services with less resource-constrained hardware for reliable and cost-effective inference maintained by dedicated teams. These LLM API Providers~\cite{artificialanalysisProviderLeaderboard} create a disjoint and inconsistent ecosystem that varies in models offered, APIs, tooling with little visibility into what drives their cost structure or how new models are added to their offering list.

Delegating infrastructure maintenance to third parties has been common practice in Web3 for years. As the resource requirements for maintaining Full Nodes~\cite{bitcoinRunningFull} increased – Solana~\cite{solanalabsSolanaValidator} recommending a minimum of 512GB of RAM for baseline functionality – the industry began relying on outsourced Remote Procedure Call (RPC) Nodes maintained by full-time DevOps teams. Today, there are dozens of major RPC providers~\cite{rpclistRPCListFind}, each servicing multiple blockchains.

\subsection{POKT Network Background}

POKT Network~\cite{poktWeb3Infrastructure} has been live on MainNet since 2020, serving hundreds of millions of daily RPC requests~\cite{poktscan} across dozens of blockchains via a heterogeneous and independent set of hardware operators. As a Decentralized Physical Infrastructure Network (DePIN), its permissionless and incentive-driven economics drive the organic addition and weeding out of supported blockchains based on customer demand and usage.

The POKT Network protocol's core Relay Mining~\cite{olshansky2023relay} algorithm acts as an on-chain metering system that cryptographically verifies how many network requests were serviced for some Application by a particular Supplier for a given service. Similar to how Bitcoin~\cite{nakamoto2008bitcoin} operates as a permissionless timestamp server, POKT serves as a permissionless, verifiable request counter or optimistic multi-tenant rate limiter. This forces Suppliers to generate a useful proof of work when servicing RPC requests, and incentivizes them to upkeep high-quality, honest services, since Applications will seek alternative providers if quality or honesty declines.

POKT Network provides this via an open internet infrastructure layer which coordinates an established network of Suppliers, atop of which a growing ecosystem of Gateways provide additional products and services. Though Applications can access the network directly, Gateways provide a mechanism to access the protocol's network while abstracting out its complexities. By vertically decoupling in this way, each network participant optimizes an aspect of performance, while preserving open access to infrastructure that is rapidly becoming a core digital public utility. 

\section{Core Problem}

\subsection{The Infrastructure Gaps}

The AI landscape is evolving rapidly and the next few years will be pivotal in determining the balance of open vs closed source foundation models, their financialization, and the API providers that facilitate access to them. Growth, adoption, and returns will be driven by tooling, incentivization and accessibility that creates equal opportunity for all of the stakeholders involved by vertically decoupling the stack and tackling the following challenges:
\begin{itemize}
    \item \textbf{Restricted model experimentation:} the resource-intensive nature of infrastructure restricts the ability of AI researchers and AI-enabled applications to explore a variety of models. Outsourcing that infrastructure to a vertically integrated partner - LLM API Providers - removes the infrastructure constraint but restricts the available range to their supported models. 
    \item \textbf{Lack of a sustainable business model for open source innovation:} independent ML engineers struggle to distribute and monetize their models and are increasingly reliant on being picked up by major infrastructure providers who, in turn, are able to squeeze their incentives. This is not conducive to sustained innovation and the emergence of supportive ecosystems.
    \item \textbf{Unequal market access:} vertically integrated infrastructure companies are incentivized to prioritize enterprise-grade customers who favour top-tier models on high-end hardware. Affordable inference for mid-tier models on mid-tier hardware, therefore, becomes harder to come by, squeezing out the middle of the market.
\end{itemize}

\subsection{POKT Network’s Unique Value Proposition}

POKT Network’s universal RPC infrastructure is perfectly primed to extend to AI Inference of in which:
An application calls a remote server, where the request acts as the prompt.
The remote server runs a procedure (the model).
The Remote Procedure Call (RPC) is completed when the generated response returns to the application.

By vertically decoupling the infrastructure layer from the product and services layer, POKT Network’s foundational infrastructure remains open and fully decentralized, while end users benefit from a growing ecosystem of Gateways that provide competitive levels of innovation, UX, and quality of service. In addition, its on-chain cryptographic rate-limiting design incentivizes high-quality service delivery and creates alignment among all network stakeholders. 

POKT Network enables Decentralized AI Inference through:
\begin{itemize}
    \item \textbf{Proven, Established Network:} Capitalizing on an established network of Suppliers that streamline access to models, along with Gateways that ensure a sufficient quality of service for end users.
    \item \textbf{Separation of Concerns:} Delineating responsibilities across the stack such that each stakeholder only focuses on their strengths to improve the efficiency of the overall ecosystem.
    \item \textbf{Incentive Alignment:} Explicit cryptographic proofs and implicit performance measurements create economic incentive alignment in a vertically decoupled ecosystem that encourages transparent competition.
    \item \textbf{Permissionless Models \& Supply:} As shown by the network’s history, its permissionless nature creates an open data and service marketplace that aggregates cost-effective hardware supply.
\end{itemize}

\section{Decentralized AI Inference Stakeholders}

A comparison of the decentralized stack versus the centralized providers is shown in figure~\ref{fig_stakeholders}. Each participant is described below (from top to bottom):

\begin{figure*}[!h]
\centering
\includegraphics[width=0.9\linewidth]{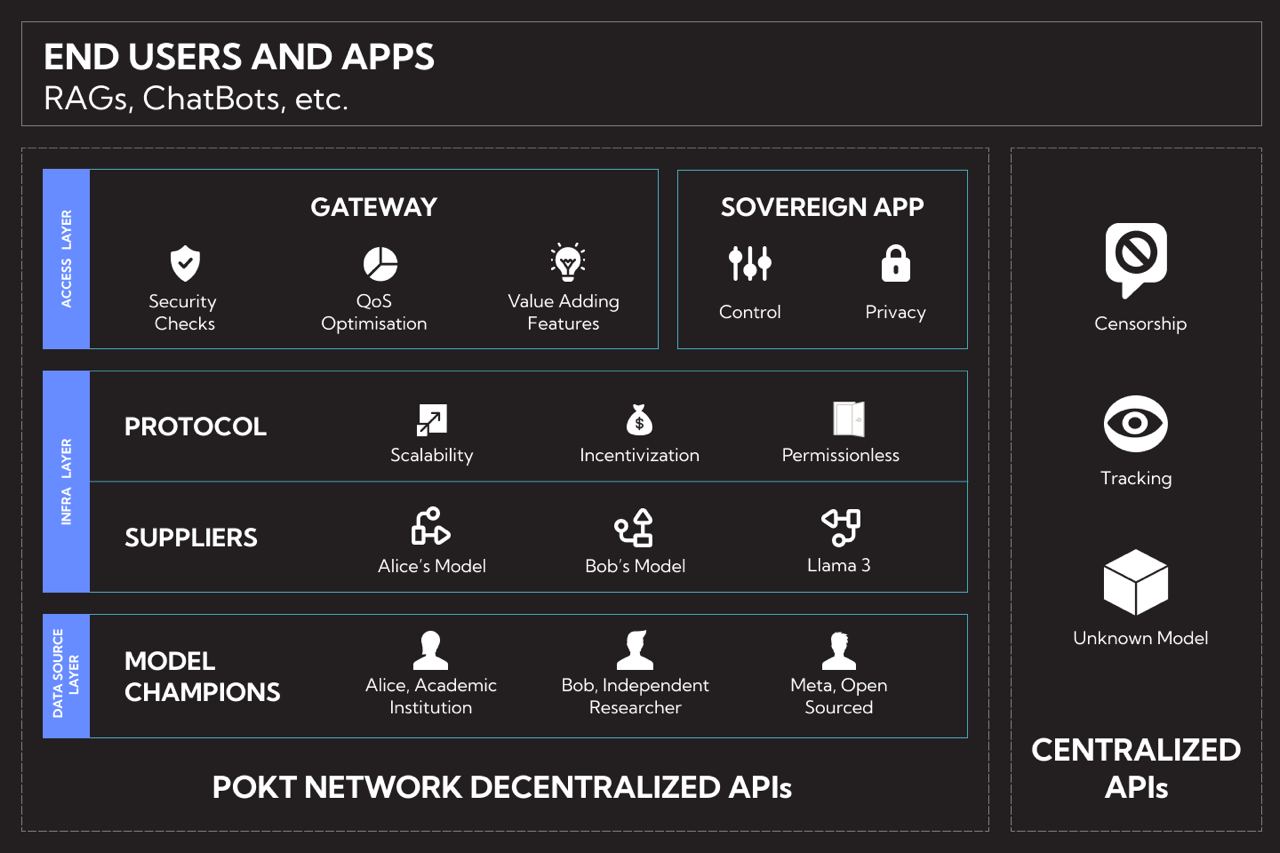}
\caption{Comparison of POKT Network's AI API actors versus centralized API service providers.}
\label{fig_stakeholders}
\end{figure*}

\subsection{Model Providers: Gateways \& Watchers}

Building and maintaining LLM infrastructure is resource intensive and likely to commoditize. As such, model Gateways are poorly incentivized to maintain it themselves, relative to dedicating the same resources to higher value-adding activities.

Gateways provide the product/services layer on top of POKT Network’s decentralized infrastructure. They serve as the entry points between Applications and the POKT Network protocol by facilitating communication and abstracting away the complexity of interacting with the protocol. They play a key part in optimizing the quality of service of the underlying infrastructure (integrity, correctness, reliability, availability, uptime, throughput, latency, security, etc) in order to provide seamless access for AI-enabled applications.

The POKT Network DAO funds and supports an open-source gateway ecosystem~\cite{poktGatewayServer} to make it as easy as possible for anyone with a strong interest in a particular model to build a business selling inference services for that model without having to build any of the underlying infrastructure themselves. The margin opportunity for Gateways comes from providing custom support, including enterprise-level Service Level Agreements (SLAs) \cite{groveSLA}, value-added features and custom pricing. POKT Network’s Gateway ecosystem provides another new and sustainable business model for open-source AI researchers to profit from their work without having to build out a globally scalable infrastructure back-end first.

Watchers are a special type of Gateways that provide checks and guarantees on the underlying Suppliers by discreetly assessing service providers while posing as regular users, ensuring they remain undetected. It offers research communities a valuable tool to assess how models perform in real-world settings, free from any biases or conflicts of interest tied to model creators or users.

\subsection{Model Users: Applications}

Most Applications will likely use Gateways to access the network, but direct access is also possible. This shifts SLA responsibility to the Application itself in exchange for full privacy and sovereignty. For instance, direct access prevents prompts from being aggregated by large Gateways. It also allows Applications to access the model marketplace directly, enabling experimentation with a potentially more diverse set of use cases. Lastly, it should result in cheaper access to services provided by the network as Applications will be able to avoid any off-chain cost structures imposed by the Gateways.

\subsection{Model Suppliers: Hardware Operators}

Suppliers are entities running inference nodes to earn POKT. They are a crucial social component of any decentralized service network, essential to both bootstrapping and growing, and are one of the most challenging assets to build.

Suppliers' core competencies include DevOps, hardware maintenance, logging, redundancy management, etc. Paired with POKT Network’s economic incentives, these abilities facilitate the rapid adoption of support for inference services across new models.

POKT Network's permissionless approach thrives on minimal friction, allowing Suppliers to stake for any inference model, and encouraging external providers to join the Supplier community. In particular, it creates opportunities to repurpose:

\begin{itemize}
    \item \textbf{Idle Inference Hardware:} Any LLM inference pipeline can be connected to POKT Network with minimal overhead. POKT Network's Relay Mining client is a coprocessor \cite{coprocessor} that acts as a lightweight "sidecar", requiring minimal resources in comparison to the hardware needed for inference.
    \item \textbf{Dormant Calculation Hardware:} Compute servers awaiting large workloads (e.g. model training) can be deployed without long-term leases or commitments. POKT Network provides a secondary revenue stream when models aren't being trained.
    \item \textbf{GPU Mining Hardware for Useful Work:} Proof of Work (PoW) operations can be repurposed to provide tailored inference models. Unlike PoW, electricity is solely used for productive tasks triggered by incoming RPC requests and paid for proportionally to the usage.
\end{itemize}

With a clear task delineation of roles and the right incentives, Suppliers focus on reducing inference costs while maximizing RPC consumption based on user demand. Developing and deploying cost-effective inference strategies, such as model quantization schemes, fast cache handling, and CPU inference, is incentivized and abstracted from the end user.

\subsection{Model Sources: Engineers \& Researchers in AI and ML}
Model Sources are individuals, teams or institutions that open-source newly trained or fine-tuned models. Often seeking users or testers, they lack the capital or expertise to deploy and manage their own performant hardware.

After publishing a model on the network, Sources leverage social forums to drive demand and collaborate with Suppliers to support it. In return, they earn a perennial revenue share from the model's success. This revenue share is a fraction of the fees paid in POKT by Applications to Suppliers for inference services, proportionate to the volume performed (i.e. the number of estimated on-chain requests).

This business model innovation enables researchers at academic institutions to earn revenue from their work's success without building customer-facing infrastructure, making it an attractive opportunity for contributors. Currently, grants and donations are the primary source of revenue supporting such stakeholders~\cite{lmsysDonationsLMSYS}.

\section{Input/Output of a Decentralized Inference Network}

\subsection{LLM Inputs to POKT Network}
The following primitives flow across the stakeholder boundaries outlined earlier. Each is necessary to enable and add value to a decentralized inference network stack.

\begin{itemize}
    \item \textbf{Open-source models:} Models supplied and accessed on the network should be open-source. Proper incentive alignment will drive adoption, solving the issue of altruistic operators common in other open-source ecosystems. Although closed-source models could be provided, this would limit availability to a small set of Suppliers.

    \item \textbf{Inference Demand:} Demand for a particular model must come from end-users or Applications. This demand can arise organically through user discovery, explicitly through model Source owners promoting their models, or via sales brought about by Gateways.

    \item \textbf{Supply Aggregation:} Supply for mid-tier models should come from commodity hardware, not specialized or scarce systems. Aggregating this supply is beneficial when targeting use cases that are too energy-intensive for end-user devices. POKT Network isn't suitable for high-end use cases requiring the latest GPUs, but it democratizes and enables supply and demand for the rest of the market.

    \item \textbf{Quality of Service Guarantees:} Operating on the network must adhere to certain SLAs. Quality can be measured via open-source frameworks, proprietary application logic (e.g., performance metrics), an on-chain reputation system, or a trusted Gateway compensated off-chain.

\end{itemize}

\subsection{LLM Outputs from POKT Network}

Despite the performance and reliability offered by centralized inference providers, they come with tradeoffs in the context of innovation, composability and privacy. For example, centralized services aggregate valuable data for future training, can unilaterally modify model performance, and can censor users at will.

A decentralized solution offers more visibility into the entire stack, benefiting the end-user or Application through the following features:

\begin{itemize}
    \item \textbf{No Downtime:} The probability of AI inference becoming unavailable is much lower since the network is inherently heterogenous, multi-tenant, multi-cloud and decentralized both geographically and geopolitically.

    \item \textbf{Model Experimentation: } Modifying the requested model type is seamless for end users, allowing requests to be routed to many providers using the same underlying protocol.

    \item \textbf{Public Model Evaluation:} Permissionless actors (Gateways, DAOs, Watchers, etc.) can build custom services to verify model performance and Supplier integrity, providing visibility and signal into actor behavior without enforcing specific attributes in the protocol on day one.

    \item \textbf{Privacy Preserving History:} The network operates as a mixing layer, where prompt inputs and inference responses are disseminated across a broader network.
    
    \item \textbf{Censorship-Free Models:}  Being permissionless and decentralized means that models aren’t subject to specific censorship, avoiding the “Woke AI”~\cite{thefpGooglesWoke} issues we’ve seen from large companies.
\end{itemize}

\section{Web3 Ecosystem Integrations}

\begin{figure*}[!h]
\centering
\includegraphics[width=0.9\linewidth]{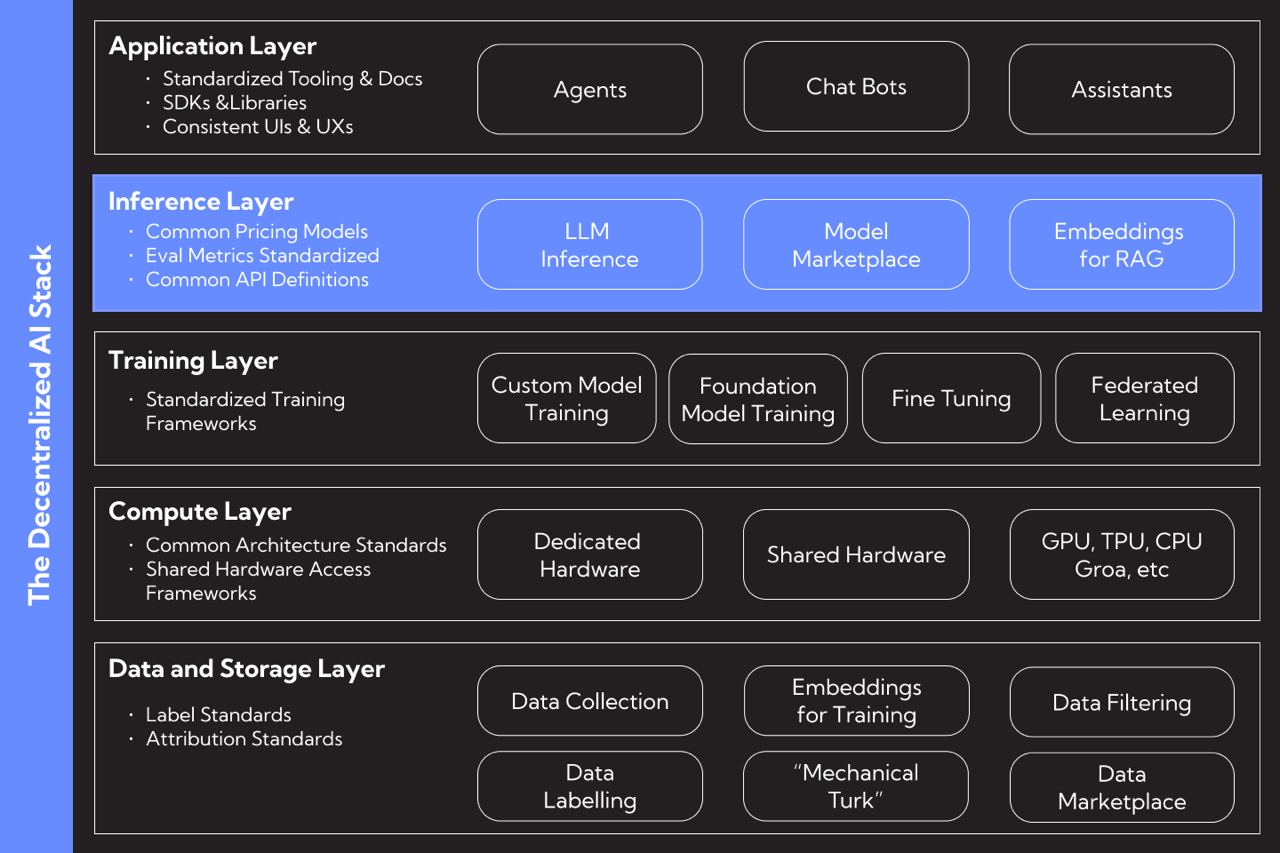}
\caption{Decentralized AI (DecAI) stack showing where POKT Network fits in the Inference Layer.}
\label{fig_stack}
\end{figure*}

POKT Network, as the largest decentralized RPC protocol for blockchain data, can integrate with other protocols in the broader Web3 ecosystem to bring additional efficiency and functionality to the Decentralized AI (DecAI) stack. The Pocket Network powers the inference layer of the ecosystem, highlighted in figure~\ref{fig_stack}.

\subsection{Data \& Storage Networks}
DecAI inferencing can leverage decentralized storage solutions like Filecoin/IPFS and Arweave, creating seamless integrations between POKT Network Suppliers and other actors in the DecAI stack.

\begin{itemize}
    \item \textbf{Network-Wide Model Storage:} AI models can be easily stored on these networks. It enhances their security, integrity, and immutability while facilitating seamless access and distribution.

    \item \textbf{Verifiable Data Storage:} Decentralized data storage enhances and incentivizes the attribution and sharability of models, labels and potentially even prompts/responses in an immutable manner.

\end{itemize}
For example, a Model owner could train their model offline, upload it to Arweave and simply point a POKT Network Supplier at the pinned file.

\subsection{Compute Networks}
DecAI inferencing can be viewed as a complementary service layer building on top of existing decentralized computing layer solutions. For example, a single POKT Node providing access to Llama 70B could act as a proxy to multiple leased GPU/TPU nodes across Akash, Render, etc to earn additional revenue when it is not used for dedicated training. DecAI inferencing can harness both dedicated or idle hardware, strengthening decentralized compute leasing networks and fostering their growth.

\subsection{Inference Networks}
DecAI inference nodes integrated into POKT Network exhibit versatility as either physical entities or logical constructs. Logical inference nodes provide abstraction and integration capabilities, sourced from diverse components such as coprocessors, agent networks, solver networks, or even inferencing blockchains.

\begin{itemize}
    \item \textbf{Flexible Deployment Model:} This abstraction enables flexible deployment and seamless integration with various DecAI infrastructural elements, accommodating different computational architectures and workflows.

    \item \textbf{Robust Ecosystem:} Physical or logical DecAI nodes contribute to a robust ecosystem supporting a wide range of inference tasks, fostering innovation and efficiency.

\end{itemize}

\subsection{Applications}
DecAI inferencing enables diverse applications, ranging from AI agents and assistants, to consumer apps and Gateways.
\begin{itemize}
    \item \textbf{AI Agents \& Assistants:} These leverage DecAI inference for personalized recommendations, task automation, and natural language understanding to enhance user experiences.

    \item \textbf{Consumer Apps:} Benefit from DecAI inference by delivering tailored content, optimizing interactions, and safeguarding data privacy.

    \item \textbf{Gateways \& IoT:} Analyze sensor data, detect anomalies, and enable autonomous edge decision-making.
\end{itemize}

\section{Conclusion}
By leveraging its established infrastructure, verifiable guarantees, and crypto-economic design, POKT Network unlocks a new infrastructure stack for open-source AI. Building on top of an ecosystem of Suppliers and Gateways that have been facilitating hundreds of millions of daily blockchain RPC requests, LLM inference stands to gain from the same reliable, performant and cost-effective services offered by the network.

POKT Network creates a utility providing bridge between open-source AI and Web3 to create new revenue streams for AI researchers without needing to maintain infrastructure or user-facing applications. In particular, AI researchers can now directly benefit from demand for their work without having to restrict access or needing to raise large amounts of capital to monetise it. While doing so, the network can leverage mid-tier idle compute resources that do not need to be exclusively leased in use-cases such as training. 

This enables POKT's stakeholders (Sources, Applications, Suppliers, and Gateways) to build innovative, sustainable, reliable, and verifiable services. Taken together, POKT Network’s approach enables a greater diversity of models to experiment with, better market access to inference infrastructure for small and medium-sized enterprises and a new sustainable business model for open-source AI researchers.

With POKT Network’s ecosystem-led approach to providing inference services, on-chain and off-chain innovation can move in parallel without being overly constrained by one another. Additionally, there are significant opportunities for integration across Web3 protocols, building the foundation for a standardized Decentralized AI stack.

This paper outlined a realistic and tangible vision that POKT Network can bring to the ecosystem in the near future. Further research and development will continue to expand the ecosystem's potential and scope, ensuring that the POKT Network remains a robust and adaptable solution within the swiftly evolving AI and blockchain landscapes.

\section{Future Work}
The intersection of POKT and AI is a broad subject, and many engaging and critical topics remain as future work. The following is a non-exhaustive list summarizing key areas of active research and development.

\begin{itemize}
    \item \textbf{Tokenomics:} Centralized services may offer superior short-term performance, but decentralized networks can accrue and provide more value as they expand. Rapid iterations in LLMs require a comprehensive tokenomics document aligning incentives with common LLM inference performance metrics~\cite{databricksInferencePerformance} like input/output token counts, Time To First Token (TTFT), Time Per Output Token (TPOT), Latency, Throughput, etc.

    \item \textbf{Trusted Execution Environment (TEE):} POKT Network can offer TEE as a marketplace option, letting users choose whether inference should be executed in a TEE. Suppliers can offer TEE via Intel SGX~\cite{intelIntelSoftware}, AMD SEV~\cite{amdsev}, AWS Nitro~\cite{amazonLightweightHypervisor}, ARM TrustZone~\cite{armTrustZoneCortexA}, etc., saving users operational overhead and earning additional revenue.

    \item \textbf{Model inference verification:} Verifying the quality and origin of model outputs is a still an active area of research, especially in a permissionless network. Our approach aims to enhance, not limit, model diversity, which introduces a new set of challenges. For example, a Supplier advertising Llama:70B may be running speculative decoding~\cite{leviathan2023fast} or even a Llama:7B to cut costs. Similar to how new strategies were developed and adopted to tackle permissionless quality-of-service as blockchain RPC volume grew, we anticipate new solutions to emerge as the network traffic for LLMs grows as well. A non-exhaustive list of approaches under active R\&D include Suppliers operating on Trusted Execution Environments (TEEs), probabilistic non-interactive on-chain quorum checks, public evaluation benchmarks by Gateways, watermarked on-chain models \cite{watermarking} or somewhat homomorphic encryption of private models \cite{cryptoeprint:2013/422}. Initially, we also anticipate "vibe based development" by Applications \cite{simonWillisonVibes} to re-route traffic to more performant actors or provide the necessary signal for Gateways that certain Suppliers may be faulty, low-quality or adversarial.
    
    \item \textbf{Adversarial Play:} Closely related to the game theory connecting tokenomics to model verification, attack mitigation requires its own dedicated document. In a permissionless environment, bad actors staking low-quality models are expected. Addressing this challenge is integral to the protocol. A separate document will be published dedicated to this topic, iterating on the tokenomics that have been driving POKT Network's sustainability over the last three plus years.

\end{itemize}

The work outlined above is a starting point and foundation layer for POKT Network's journey into DecAI. This list doesn't constrain the protocol's future, as POKT Network will continue exploring opportunities and refining the roadmap.

\ifCLASSOPTIONcompsoc
  \section*{Acknowledgments}
\else
  \section*{Acknowledgment}
\fi

The authors would like to thank Adrienne and Dermot from the POKT Network Foundation for their guidance and graphics construction and the whole POKT Network community for the feedback that helped to shape this document.



\bibliographystyle{IEEEtran}
\bibliography{refs}
%



\end{document}